\documentstyle[12pt,
epsfig]{article}
\begin{document}
\newcommand{\tcr}{T_{cr}}
\newcommand{\sxj}{\sigma_j}
\newcommand{\vt}{\tilde{v}}
\newcommand{\sxjt}{\tilde{\sigma_j}}
\newcommand{\dtt}{\tilde{\delta}}
\newcommand{\Lt}{\tilde{\Lambda}}
\newcommand{\Ut}{\tilde{U}}
\newcommand{\dj}{\delta j}
\newcommand{\du}{\delta u}
\newcommand{\Vt}{\tilde{V}}
\newcommand{\Ct}{\tilde{C}}
\newcommand{\dm}{\delta \mu}
\newcommand{\dms}{\delta m^2}
\newcommand{\dmst}{\delta \tilde{m}^2}
\newcommand{\dlx}{\delta \lambda}
\newcommand{\phit}{\tilde{\phi}}
\newcommand{\sxt}{\tilde{\sigma}}
\newcommand{\sxb}{\bar{\sigma}}
\newcommand{\lxb}{\bar{\lambda}}
\newcommand{\gb}{\bar{g}}
\newcommand{\rht}{\tilde{\rho}}
\newcommand{\jt}{\tilde{j}}
\newcommand{\djt}{\delta \tilde{j}}
\newcommand{\done}{\delta\phi_1}
\newcommand{\dy}{\delta y}
\newcommand{\tr}{\rm Tr}
\newcommand{\sx}{\sigma}
\newcommand{\mpl}{m_{Pl}}
\newcommand{\Mpl}{M_{Pl}}
\newcommand{\lx}{\lambda}
\newcommand{\Lx}{\Lambda}
\newcommand{\kx}{\kappa}
\newcommand{\ex}{\epsilon}
\newcommand{\be}{\begin{equation}}
\newcommand{\ee}{\end{equation}}
\newcommand{\eesn}{\end{subequations}}
\newcommand{\besn}{\begin{subequations}}
\newcommand{\beq}{\begin{eqalignno}}
\newcommand{\eeq}{\end{eqalignno}}
\def \lta {\mathrel{\vcenter
     {\hbox{$<$}\nointerlineskip\hbox{$\sim$}}}}
\def \gta {\mathrel{\vcenter
     {\hbox{$>$}\nointerlineskip\hbox{$\sim$}}}}

\newcommand{\nwc}{\newcommand}
%
%
\nwc{\cl}  {\clubsuit}
\nwc{\di}  {\diamondsuit}
\nwc{\sps} {\spadesuit}
\nwc{\hyp} {\hyphenation}
\nwc{\ba}  {\begin{array}}
\nwc{\ea}  {\end{array}}
\nwc{\bdm} {\begin{displaymath}}
\nwc{\edm} {\end{displaymath}}
\nwc{\bea} {\be\ba{rcl}}
\nwc{\eea} {\ea\ee}
\nwc{\ben} {\begin{eqnarray}}
\nwc{\een} {\end{eqnarray}}
\nwc{\bda} {\bdm\ba{lcl}}
\nwc{\eda} {\ea\edm}
\nwc{\bc}  {\begin{center}}
\nwc{\ec}  {\end{center}}
\nwc{\ds}  {\displaystyle}
\nwc{\bmat}{\left(\ba}
\nwc{\emat}{\ea\right)}
\nwc{\non} {\nonumber}
\nwc{\bib} {\bibitem}
\nwc{\lra} {\longrightarrow}
\nwc{\Llra}{\Longleftrightarrow}
\nwc{\ra}  {\rightarrow}
\nwc{\Ra}  {\Rightarrow}
\nwc{\lmt} {\longmapsto}
\nwc{\pa} {\partial}
\nwc{\iy}  {\infty}
\nwc{\ovl}  {\overline}
\nwc{\hm}  {\hspace{3mm}}
\nwc{\lf}  {\left}
\nwc{\ri}  {\right}
\nwc{\lm}  {\limits}
\nwc{\lb}  {\lbrack}
\nwc{\rb}  {\rbrack}
\nwc{\ov}  {\over}
\nwc{\pr}  {\prime}
\nwc{\nnn} {\nonumber \vspace{.2cm} \\ }
\nwc{\Sc}  {{\cal S}}
\nwc{\Lc}  {{\cal L}}
\nwc{\Rc}  {{\cal R}}
\nwc{\Dc}  {{\cal D}}
\nwc{\Oc}  {{\cal O}}
\nwc{\Cc}  {{\cal C}}
\nwc{\Pc}  {{\cal P}}
\nwc{\Mc}  {{\cal M}}
\nwc{\Ec}  {{\cal E}}
\nwc{\Fc}  {{\cal F}}
\nwc{\Hc}  {{\cal H}}
\nwc{\Kc}  {{\cal K}}
\nwc{\Xc}  {{\cal X}}
\nwc{\Gc}  {{\cal G}}
\nwc{\Zc}  {{\cal Z}}
\nwc{\Nc}  {{\cal N}}
\nwc{\fca} {{\cal f}}
\nwc{\xc}  {{\cal x}}
\nwc{\Ac}  {{\cal A}}
\nwc{\Bc}  {{\cal B}}
\nwc{\Uc}  {{\cal U}}
\nwc{\Vc}  {{\cal V}}
%
%
\nwc{\Th} {\Theta}
\nwc{\th} {\theta}
\nwc{\vth} {\vartheta}
\nwc{\eps}{\epsilon}
\nwc{\si} {\sigma}
\nwc{\Gm} {\Gamma}
\nwc{\gm} {\gamma}
\nwc{\bt} {\beta}
\nwc{\La} {\Lambda}
\nwc{\la} {\lambda}
\nwc{\om} {\omega}
\nwc{\Om} {\Omega}
\nwc{\dt} {\delta T}
\nwc{\Si} {\Sigma}
\nwc{\Dt} {\Delta}
\nwc{\al} {\alpha}
\nwc{\vph}{\varphi}
\nwc{\zt} {\zeta}
%
%
\def\tr{\mathop{\rm tr}}
\def\Tr{\mathop{\rm Tr}}
\def\Det{\mathop{\rm Det}}
\def\Im{\mathop{\rm Im}}
\def\Re{\mathop{\rm Re}}
\def\secder#1#2#3{{\partial^2 #1\over\partial #2 \partial #3}}
\def\bra#1{\left\langle #1\right|}
\def\ket#1{\left| #1\right\rangle}
\def\VEV#1{\left\langle #1\right\rangle}
\def\gdot#1{\rlap{$#1$}/}
\def\abs#1{\left| #1\right|}
\def\pr#1{#1^\prime}
\def\ltap{\raisebox{-.4ex}{\rlap{$\sim$}} \raisebox{.4ex}{$<$}}
\def\gtap{\raisebox{-.4ex}{\rlap{$\sim$}} \raisebox{.4ex}{$>$}}
\nwc{\Id}  {{\bf 1}}
\nwc{\diag} {{\rm diag}}
\nwc{\inv}  {{\rm inv}}
\nwc{\mod}  {{\rm mod}}
\nwc{\hal} {\frac{1}{2}}
\nwc{\tpi}  {2\pi i}
\def\contract{\makebox[1.2em][c]{
        \mbox{\rule{.6em}{.01truein}\rule{.01truein}{.6em}}}}
\def\slash#1{#1\!\!\!/\!\,\,}
%
%
\def\KK{{\rm I\kern -.2em  K}}
\def\NN{{\rm I\kern -.16em N}}
\def\RR{{\rm I\kern -.2em  R}}
\def\ZZ{Z \kern -.43em Z}
\def\QQ{{\rm \kern .25em
             \vrule height1.4ex depth-.12ex width.06em\kern-.31em Q}}
\def\CC{{\rm \kern .25em
             \vrule height1.4ex depth-.12ex width.06em\kern-.31em C}}
\def\ZZZ{Z\kern -0.31em Z}

\def \Msol {M_\odot}
\def\eV {\,{\rm  eV}}     
\def\KeV {\,{\rm  KeV}}     
\def\MeV {\,{\rm  MeV}}
\def\GeV {\,{\rm  GeV}}     
\def\TeV {\,{\rm  TeV}}     
\def\fm {\,{\rm  fm}}

\def\ap#1{Annals of Physics {\bf #1}}
\def\cmp#1{Comm. Math. Phys. {\bf #1}}
\def\hpa#1{Helv. Phys. Acta {\bf #1}}
\def\ijmpa#1{Int. J. Mod. Phys. {\bf A#1}}
\def\jpc#1{J. Phys. {\bf C#1}}
\def\mpla#1{Mod. Phys. Lett. {\bf A#1}}
\def\npa#1{Nucl. Phys. {\bf A#1}}
\def\npb#1{Nucl. Phys. {\bf B#1}}
\def\nc#1{Nuovo Cim. {\bf #1}}
\def\pha#1{Physica {\bf A#1}}
\def\pla#1{Phys. Lett. {\bf #1A}}
\def\plb#1{Phys. Lett. {\bf #1B}}
\def\pr#1{Phys. Rev. {\bf #1}}
\def\pra#1{Phys. Rev. {\bf A#1 }}
\def\prb#1{Phys. Rev. {\bf B#1 }}
\def\prp#1{Phys. Rep. {\bf #1}}
\def\prc#1{Phys. Rep. {\bf C#1}}
\def\prd#1{Phys. Rev. {\bf D#1 }}
\def\ptp#1{Progr. Theor. Phys. {\bf #1}}
\def\rmp#1{Rev. Mod. Phys. {\bf #1}}
\def\rnc#1{Riv. Nuo. Cim. {\bf #1}}
\def\zpc#1{Z. Phys. {\bf C#1}}
\def\APP#1{Acta Phys.~Pol.~{\bf #1}}
\def\AP#1{Annals of Physics~{\bf #1}}
\def\CMP#1{Comm. Math. Phys.~{\bf #1}}
\def\CNPP#1{Comm. Nucl. Part. Phys.~{\bf #1}}
\def\HPA#1{Helv. Phys. Acta~{\bf #1}}
\def\IJMP#1{Int. J. Mod. Phys.~{\bf #1}}
\def\JP#1{J. Phys.~{\bf #1}}
\def\MPL#1{Mod. Phys. Lett.~{\bf #1}}
\def\NP#1{Nucl. Phys.~{\bf #1}}
\def\NPPS#1{Nucl. Phys. Proc. Suppl.~{\bf #1}}
\def\NC#1{Nuovo Cim.~{\bf #1}}
\def\PH#1{Physica {\bf #1}}
\def\PL#1{Phys. Lett.~{\bf #1}}
\def\PR#1{Phys. Rev.~{\bf #1}}
\def\PRP#1{Phys. Rep.~{\bf #1}}
\def\PRL#1{Phys. Rev. Lett.~{\bf #1}}
\def\PNAS#1{Proc. Nat. Acad. Sc.~{\bf #1}}
\def\PTP#1{Progr. Theor. Phys.~{\bf #1}}
\def\RMP#1{Rev. Mod. Phys.~{\bf #1}}
\def\RNC#1{Riv. Nuo. Cim.~{\bf #1}}
\def\ZP#1{Z. Phys.~{\bf #1}}

\pagestyle{empty}
\noindent
\begin{flushright}
January 2004
\\
\end{flushright} 
\vspace{3cm}
\begin{center}
{ \Large \bf
The Universal Equation of State \\
near \\ the Critical Point of QCD \\
} 
\vspace{1.5cm}
{\Large 
N. Brouzakis  and N. Tetradis 
} 
\\
\vspace{0.5cm}
{\it
Department of Physics, University of Athens,
Zographou 157 71, Greece
} 
\\
\vspace{3cm}
\abstract{
We study the universal properties of the phase diagram of QCD
near the critical point using the exact renormalization
group. For two-flavour QCD and zero quark masses we derive the 
universal equation of state in the vicinity of the tricritical point.
For non-zero quark masses 
we explain how the universal equation of state
of the Ising universality class can be used in order to describe 
the physical behaviour near the line of critical points.
The effective exponents that
parametrize the growth of physical quantities, such as the correlation
length, are given by combinations of the critical exponents of the
Ising class that depend on the path along which the critical point
is approached. 
In general the critical region, in which such quantities become large,  
is smaller than naively expected. 

} 
\end{center}

\newpage

\pagestyle{plain}

\setcounter{equation}{0}

\section{Introduction}

The most prominent feature of the phase diagram of QCD 
at non-zero temperature and baryonic density
is the critical point that marks the end of the
line of first-order phase transitions. 
(For a review see ref. \cite{Rajagopal:2000wf}.) 
Its exact location and the size of the critical region around it 
determine its relevance for the heavy-ion
collision experiments at RHIC and LHC. 

There exist several analytical studies of the phase diagram near the critical
point \cite{Pisarski:ms}--\cite{Kiriyama:2000yp}. Lattice 
simulations are a source of precise information for the details of
the phase diagram. Recent studies 
\cite{Fodor:2001pe,deForcrand:2002ci,Allton:2002zi} have overcome
the difficulties associated with simulating systems with
non-zero chemical potential, even though the continuum limit or realistic 
quark masses 
have not been reached yet.

The critical equation of state encodes all the physical information 
for a system near a critical point in the presence of an external
source. In the case of QCD, the parameters that can be adjusted in
order to approach the critical point are the temperature $T$ and baryonic
chemical potential $\mu$. For two-flavour 
QCD, the role of the external source $j$ is 
played by the current quark mass that breaks explicitly 
the chiral symmetry of the QCD Lagrangian.
If isospin-breaking effects are neglected, the source is 
a (linear) function of the common mass of the light quarks. 

An analytical description of
the critical region requires the use of the renormalization group. 
The study of first-order phase transitions
is easier within the Wilsonian (or exact) formulation \cite{Wilson:1973jj}. 
We shall use the formalism of the effective average action
$\Gamma_k$ 
\cite{Wetterich:1989xg,rome}, that results by integrating
out (quantum or thermal) fluctuations above a certain cutoff $k$.
The dependence of $\Gamma_k$ on 
$k$ is given by an exact flow equation
\cite{Wetterich:yh,Tetradis:1993ts}.
When $k$ is lowered to zero $\Gamma_k$
becomes the effective action.
The effective average action 
can by used in order to desribe very efficiently the universal and
non-universal aspects of second-order phase transitions, as well as
strong or weak first-order phase transitions \cite{Berges:2000ew}.

In order to discuss the phase diagram we have to
employ an appropriate order parameter. The usual choice is the 
quark-antiquark condensate that is associated with the spontaneous
breaking of the chiral symmetry of the QCD Lagrangian.
For two flavours
an equivalent description uses as effective degress of freedom
four mesonic scalar
fields (the $\sigma$-field and the three pions $\pi_i$),
arranged in 
a $2\times 2$ matrix
\begin{equation}
  \Phi=\frac{1}{2}\left(\si+i\vec{\pi}\cdot\vec{\tau}\right),
  \label{phi}
\end{equation}
where $\tau_i$ $(i=1,2,3)$ denote the Pauli matrices. 
The various interactions must be
invariant under a global $SU(2)_L \times SU(2)_R$
symmetry acting on $\Phi$.
If the $\sigma$-field (that corresponds to 
a condensate ${\bar u}u+{\bar d}d$) develops an expectation value the
symmetry is broken down to $SU(2)_{L+R}$. 
The explicit breaking of the chiral symmetry through the
current 
quark masses can be incorporated as well. It corresponds to the interaction
with an external source through a term $-j \sigma$ in the Lagrangian.

The full QCD dynamics cannot be described by an effective 
Lagrangian involving only mesonic fields. More complicated effective
descriptions, such as 
the linear quark-meson model, have been used for the discussion of
the phase diagram for low chemical potential
\cite{Jungnickel:1995fp,Berges:1998sd,Berges:1997eu}.
In ref. \cite{Tetradis:2003qa} it was shown that the expected 
structure of the phase diagram for two-flavour QCD 
can be reproduced within this model. In particular, 
in the absence of an external source (zero current 
quark mass), the phase diagram
contains a line of first-order phase transitions and a line of 
second-order ones, that meet at a tricritical point. The second-order
phase transitions belong to the $O(4)$ universality class, while the
tricritical point is associated with the Gaussian fixed point and is
described by mean-field theory.
In the presence of
a source the  first-order line ends at a critical point. The second
order phase-transition at this point belongs to the Ising universality 
class. The reason is that all degrees of freedom other than the $\sx$-field
remain massive and decouple at low energy scales. For varying external
source we obtain a line of second-order phase transitions, that starts
at the tricritical point and marks the boundary of a surface of
first-order transitions.

The universal properties associated with the lines of second-order
phase transitions in the phase diagram do not depend on the effective
model employed for the description of the physical system (the mesonic
model described above, a discretized version of the QCD Langrangian employed
in lattice simulations, etc). The crucial ingredient is the inclusion
of the degrees of freedom that become critical (massless) at a transition,
and the symmetries that characterize the effective interactions
\cite{Toldin:2003hq}.
For the same reason the inclusion of the more massive quarks is not expected
to modify the universal behaviour. It may only shift the location of the
various lines on the phase diagram.

The purpose of this paper is to derive the universal equation of state
of QCD in the region of the phase diagram 
near the line of critical points. We concentrate on the dependence on the
quantities that can be varied 
In the following section we
desribe a simplified model that has a phase diagram similar to that
of QCD at relatively small values of the chemical potential. In section 3
we discuss the phase diagram in the context of mean-field theory. In section 4
we describe the renormalization-group approach. We also determine the initial
conditions for the renormalization-group flow in the context of the 
quark-meson model.
In section 5 we derive the
equation of state in the absence of an external source. In section 6 we
derive the equation of state for a non-zero external source in the vicinity
of the line of critical points. In the final section 7 we discuss the
implications of our results for the possibility of observing the universal
critical behaviour of QCD in heavy-ion experiments.

\section{The model}

As we are interested only in the universal characteristics near the
second-order phase transitions, we need only discuss a simplified model
that preserves the basic structure of the phase diagram of two-flavour QCD. 
We consider a Lagrangian in three dimensions
that contains standard kinetic terms for the fields 
$\sx$, $\pi_i$  ($i=1,2,3$), and the potential
\be
V(\rho)=m^2\rho+\frac{1}{2}\lx\rho^2+\frac{1}{3}g\rho^3,
\label{mpot} \ee
with $\rho=\Tr\left(\Phi^\dagger \Phi\right)=(\sx^2+\pi_i\pi^i)/2$. 
The couplings $m^2$, $\lx$, $g$ are considered
functions of the temperature $T$ and the chemical potential $\mu$. 
We assume that $g$ is always positive, so that the potential is
bounded from below.

In ref. \cite{Tetradis:2003qa} it was shown that such a potential
emerges in the quark-meson model after the integration of the
temperature fluctuations with non-zero Matsubara frequences. 
The fermions are completely integrated out as their lowest Matsubara
frequency is non-zero. The only mode that survives at energy scales
below the temperature is the zero-mode of the mesonic field. 
Thus the theory becomes effectively three-dimensional. 
The effective potential around the origin has a structure similar to
that of eq. (\ref{mpot}). In section 4 we calculate the effective
potential in the context of an effective model of QCD, the linear 
quark-meson model.

The various fixed points that determine the structure of the
phase diagram appear during the integration of the low-energy 
fluctuations of the zero mode. Theories for which the renormalization-group
flow is dominated by an infrared fixed point become insensitive to the 
details of the ultraviolet formulation.
For this reason the simplified model we are considering is sufficient for
the discussion of the universal properties 
of the phase diagram of QCD near the lines of second-order phase transitions.
 
The effect of a current quark mass can be taken into account by adding a
term $-j \sx$, with $j$ a (linear) function of the mass.

\section{Mean-field theory}

It is instructive to discuss the phase diagram of our simplified 
model neglecting the field fluctuations. We allow for variations of
$m^2$ and $\lx$, while we assume that the coupling $g$ remains constant.
This is a good approximation of the behaviour of the potential studied
in ref. \cite{Tetradis:2003qa}.

We begin by studying the phase diagram in the absence of an external source
($j=0$).
Let us consider first the case $\lx>0$. For
$m^2>0$ the minimum of the potential is located at $\sx=\pi_i=0$ and
the system is invariant under an $O(4)$ symmetry. 
For $m^2<0$ the minimum moves away from the origin. Without loss of
generality we take it along the $\sx$ axis. The symmetry is broken down
to $O(3)$. The $\pi$'s, which play the role of the
Goldstone fields, become massless at the minimum. 
If $m^2(T,\mu)$ has a zero at a certain value $T=T_{cr}$, while 
$\lx(T_{cr},\mu)>0$, the system undergoes a second-order 
phase transition at this point.
The minimum of the potential 
behaves as $\sx_0\sim |T-T_{cr}|^{1/2}$ slightly below the critical 
temperature. The critical exponent $\beta$ takes the mean-field value
$\beta=1/2$.

Let us consider now the case $\lx<0$. 
It is easy to check that, as $m^2$ increases
from negative to positive values (through an increase of $T$ for example), 
the system undergoes a first-order
phase transition. For $|\lx|$ approaching zero the phase transition
becomes progressively weaker: The discontinuity in the order parameter 
(the value of $\sx$ at the minimum) approaches zero. For $\lx=0$ the
phase transition becomes second order. The minimum of the potential behaves
as $\sx_0\sim |T-T_{cr}|^{1/4}$. The critical exponent $\beta$ takes 
the value $\beta=1/4$ for this particular point.

Let us assume now for simplicity 
that $\lx$ is a decreasing function of $\mu$ only, 
and has a zero at $\mu=\mu_*$.
If we consider the phase transitions for increasing $T$ and fixed 
$\mu$ we find a line of second-order phase transitions for $\mu < \mu_*$,
and a line of first-order transitions for $\mu > \mu_*$.
The two lines meet at the
special point $(T_*,\mu_*)$, where $T_*=T_{cr}$ for $\mu=\mu_*$. This
point is characterized as a tricritical point. (For a review of the
theory of tricritical points see ref. \cite{lawrie}.)
In the general case $\lx$ will be a decreasing function of a linear combination
of $\mu$ and $T$. The structure of the phase diagram remains the same. 
Simply there is a linear combination of $T$ and $\mu$ that generates 
displacements along the lines of first- and second-order phase transitions
near the tricritical point, and a different one that moves the system
through the phase transition.

If a source term $-j \sx$ is added to the potential of eq. (\ref{mpot})
the $O(4)$ symmetry is explicitly broken. The phase diagram is modified
significantly even for small $j$.
The second-order phase transitions, observed for 
$\lx >0$, disappear. The reason is that the minimum of the potential
is always at a value $\sx \not= 0$ that moves close to zero for increasing
$T$. For small $j$ the mass term at the minimum approaches zero at 
a value of $T$ near what we defined as $T_{cr}$ for $j=0$. However,
no genuine phase transition appears. Instead we observe an analytical
crossover. 

\begin{figure}[t]
 \centerline{\epsfig{figure=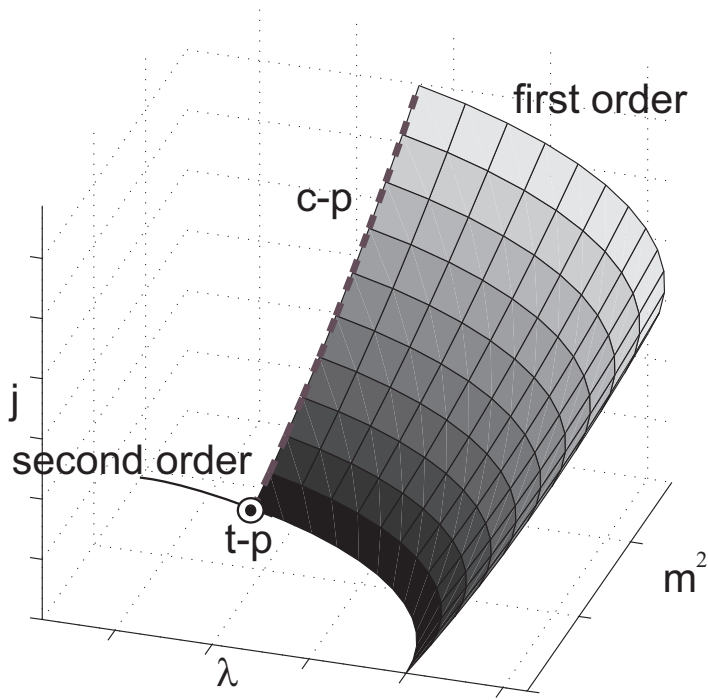,width=13cm,angle=0}}
 \caption{\it
The phase diagram.
}
 \label{fig0}
 \end{figure}

The line of first-order transitions 
persists for $j\not= 0$. However, 
at the critical temperature 
the minimum jumps discontinuously between two non-zero
values of $\sx$ and never becomes zero. The line ends at a new special
point, whose nature can be examined by considering the $\sx$-derivative of
the potential $V(\rho)$ of eq. (\ref{mpot}). 
For a first-order phase transition to occur, 
$\partial V/\partial \sx$ must become equal to $j$ for three non-zero
values of $\sx$. This requires $\partial^2 V/\partial \sx^2 =0$
at two values of $\sx$. The critical point corresponds to the situation that
all these values merge to one point.  At the minimum $\sx_*$ 
of the potential $\Vt(\sx;j)=V(\sx,\pi_i=0)-j\sx$ 
at the critical point, we have:
$d\Vt/d\sx=d^2\Vt/d\sx^2 =d^3\Vt/d\sx^3=0$. 
It can be checked that this requires
$\lx <0$ and can be achieved by fine-tuning
$m^2$ and $\lx$ for given $j$. The minimum $\sx_*$ is then completely
determined.
At the critical point we expect
a second-order phase transition for the deviation of $\sx$ from $\sx_*$.
As $d^4\Vt/d\sx^4\not= 0$ at the critical point,  
the critical exponent $\beta$ takes the value $\beta=1/2$.
The $\pi$'s are massive, because
$m^2_\pi=\partial^2 V(\rho)/\partial \pi_i^2 = dV(\rho)/d\rho=j/\sx_*$ at the 
critical point.

The phase diagram we discussed is presented in fig. \ref{fig0}.
If we take into account
the effect of fluctuations of the fields, the potential will have
a more complicated form than the simple assumption of eq. (\ref{mpot}).
As we shall see the qualitative structure of the phase diagram is not
modified. 
However, the nature of the
fixed points may differ from the predictions of mean-field theory.

\section{The renormalization-group flow}

The most efficient way to study a physical
system near a phase transition is through the effective potential.
In the formulation of the renormalization group which we are employing, the
dependence of the potential of a three-dimensional theory
on a ``coarse-graining'' scale $k$
is described by the equation
\cite{Wetterich:yh,Berges:2000ew}
\begin{equation}
\frac{\partial}{\partial t} u_k(\sxt)=-3 u_k + \frac{1}{2}(1+\eta)
\sxt \, u'_k+
\frac{1}{4\pi^2} \left[ 
l^3_0\left( u''_k\right) 
+3l^3_0\left( u'_k/\sxt \right) 
\right],
\label{flow} \end{equation}
where $t=\ln(k/\Lx)$ and we have defined the dimensionless quantities
\be
u_k=k^{-3} U_k, ~~~~~~~~~~
\sxt= k^{-\frac{1}{2}} Z_k^{\frac{1}{2}} \sx.
\label{scalvar} \ee
Primes denote derivatives with respect to $\sxt$.
The scale-dependent potential $U_k$ results from the integration of
the field fluctuations with characteristic momenta larger than $k$. In the
limit $k\to 0$ it becomes equal to the effective potential.
It is possible that the three-dimensional theory described by the potential
$U_k$ has its origin in the dimensional reduction of a 
more complicated, higher-dimensional theory. This situation arises during
the study of the phase diagram of QCD at non-zero temperature $T$ and
baryonic chemical potential $\mu$ (see below).

The above equation can be derived from an exact flow equation
if only two terms are retained in the effective action:
the potential and a
kinetic term that includes a $\Phi$-independent 
wavefunction renormalization $Z_k$ 
\cite{Berges:2000ew}. 
The anomalous dimension $\eta=-d(\ln Z_k)/dt$ describes the
scale dependence of $Z_k$, and can be determined starting from
the exact flow equation.
It becomes constant when the flow approaches
a fixed point \cite{Berges:2000ew}.

The ``threshold'' function $l^3_0$
is a particular case of
\be
l^d_n(w) = 
 \frac{n}{2} k^{2n-d+1} \pi^{-\frac{d}{2}}\Gamma \left( \frac{d}{2} \right) 
\int d^d \vec{q}~~ \frac{\partial P_{k}}{ \partial k} 
(P_{k} + w\,k^2)^{-(n+1)}.
\label{fourfive} \ee
The inverse propagator for massless fields
$P_{k}$ has been modified so at to forbid the propagation of
modes with momenta below the ``coarse-graining'' scale $k$.
Several forms of the modified propagator are possible, which
correspond to sharper or smoother 
cutoffs of the low-momentum modes \cite{Berges:2000ew}.

The continuous reduction of $k$, from the ultraviolet scale $\Lx$
to zero, permits the gradual integration of fluctuations 
with different characteristic momenta.
The function $l^3_0(w)$ falls off 
for large values of $w$ following a power law. As 
a result it introduces a threshold behaviour for the 
contributions of massive modes to the evolution equation. 
When the running squared mass of a massive mode 
becomes much larger than the scale $k^2$, 
the mode decouples and the respective contribution vanishes.
In eq. (\ref{flow}) we distinguish contributions from two
different types of fields: the radial mode (the $\sx$-field)
and the three Goldstone modes (the pions). Their masses are expressed
through the derivatives of the effective potential. 

The initial condition that is needed for the solution of the
evolution equation (\ref{flow}) is given by eq. (\ref{mpot}).
The potential at the ultraviolet scale $\Lx$ is identified with 
the classical (or bare) potential: $U_\Lx =V$. The integration
of field fluctuation with momenta between $\Lx$ and zero leads the
determination of the effective potential: $U_{eff}=U_0\equiv U$.

It is possible that the theory at $k=\Lx$ results from the integration of
some complicated ultraviolet theory. This situation arises during
the study of the phase diagram of QCD at non-zero temperature $T$ and
baryonic chemical potential $\mu$
\cite{Rajagopal:2000wf,Berges:1998sd,Berges:1997eu,
Tetradis:2003qa}. An effective theory that captures many of the features of
QCD is the quark-meson model \cite{Jungnickel:1995fp}. It employs  
mesons and quarks as physical degrees of freedom. The QCD phase transition
is associated with the breaking
of the chiral symmetry through a non-zero expectation value for
the mesonic $\sx$-field. The connection with the model we are studying 
is made through the phenomenon of dimensional reduction at non-zero 
temperature. At energy scales below the temperature 
a four-dimensional theory can be described in terms of effective 
three-dimensional degrees of freedom. In a Fourier decomposition of 
a scalar field only the term with zero Matsubara frequency (the zero mode)
survives. The other modes decouple, as they get a mass proportional to 
the temperature. The fermions do not have a zero mode and they decouple
completely.

In the context of the quark-meson model with two flavours an analytical 
expression has been derived for the form of the potential at a scale 
$k=\Lx=T$ \cite{Tetradis:2003qa}. It is 
\be
U'_{\Lx}(\rho;T,\mu) \simeq  \lx_{0R} \left[
\rho-\rho_{0R}-\frac{3}{4\pi^2} T^2 
\left( 
\frac{\sqrt{\pi}}{\theta_2} -\frac{\pi^2}{3} \right)
\right] +I_F(\rho;T,\mu),
\label{init} \ee
where
\begin{eqnarray}
I_F(\rho;T,\mu)&=&
-N_c\frac{h^4}{16\pi^2}\rho \ln \left( 
\frac{\frac{h^2\rho}{2\Lt^2}}{1+\frac{h^2\rho}{2\Lt^2}}\right)
+{V'}_{F}(\rho;T,\mu)+{V'}_{F}(\rho;T,-\mu) \, ,
\label{if} \\
   \ds{{V}_{F}(\rho;T,\mu)} &=&
    \ds{
    -4 N_c T 
    \int\limits_{-\infty}^{\infty}
    \frac{d^3\vec{q}}{(2\pi)^3} 
    \ln \left(1 + \exp 
    \left[- \frac{1}{T}
     \left(\sqrt{\vec{q}^{\,2}+h^2 \rho/2}-\mu\right)\right]
    \right) 
     }
     \, .
\label{f23}  
\end{eqnarray}
Here $\rho_{0R}$ is a renormalized expectation value for the mesonic
field at zero temperature and chemical
potential, while $\lx_{0R}$ is the renormalized mesonic quartic coupling. 
The term $\sim T^2$ in eq. (\ref{init}) results from the integration of
the scalar thermal fluctuations with non-zero Matsubara frequences.
The effect of the
fermionic quantum and thermal fluctuations is incorporated in the
quantity $I_F(\rho;T,\mu)$.
Even at zero temperature,
the fermionic quantum fluctuations give a contribution
(the first term in the 
r.h.s. of eq. (\ref{if})) that has to be taken into account when defining
the vacuum of the zero-temperature theory.
The quantity ${V}_{F}(\rho;T,\mu)$ is the free energy of a 
fermionic gas at non-zero temperature and chemical potential. 
We have taken into account only two quark flavours (corresponding to the
light $u$- and $d$-quarks), as well as their antiparticles
through ${V}_{F}(\rho;T,-\mu)$. The parameter $N_c=3$ stands for
the number of colours. The scale $\Lt$ corresponds to the energy scale at which
the quark-meson model becomes an effective description of QCD. It is close to
650 GeV \cite{Jungnickel:1995fp}, while $T \simeq 200$ MeV.
The constituent quark mass is given by $M^2_q=h^2 \rho/2$.

We emphasize that the above expressions have been derived for small
$\lx_{0R}$ and $h$. If we would like to make contact with 
realistic mesonic physics we must 
reproduce the correct pion decay constant $f_\pi$, the
$\sigma$-field mass and the constituent quark mass at zero temperature and
chemical potential.
If we ignore the logarithmic contribution in $I_F(\rho;T,\mu)$,
we have to set
$\sqrt{2\rho_{0R}}\simeq f_\pi=87$ MeV, 
$m_\sx\simeq\sqrt{2\lx_{0R}\rho_{0R}}\simeq 600$ MeV, 
$M_q=\sqrt{h^2\rho_{0R}/2} \simeq 300$ MeV.
This leads to $\lx,h={\cal O}(10).$
For this reason the essentially perturbative expressions
(\ref{init})--(\ref{f23}) must be used only for 
qualitative discussions of the behaviour
of realistic QCD. Whenever we make use of these expressions 
in the following we assume that $\lx_{0R}$ and $h$ are small, consistently
with their derivation. 

The quantity $I_F(\rho;T,\mu)$ can be expanded in powers of
$\rho$ as
\begin{eqnarray}
I_F(\rho;T,\mu) &=&
\frac{h^2N_c}{12}\left(T^2+\frac{3}{\pi^2}\mu^2 \right)
+\frac{h^4 N_c}{16\pi^2}
\Biggl[
-1+2\gamma_e+\ln\left(\frac{\Lt^2}{4T^2}\right) \Biggr.
\nonumber \\
&&~
+ 2\,{\rm Li}^{(1,0)}\left(0,-\exp\left(\frac{\mu}{T} \right) \right)
\Biggl. +2\,{\rm Li}^{(1,0)}\left(0,-\exp\left(-\frac{\mu}{T} \right) \right)
\Biggr]\rho
+{\cal O}(\rho^2),
\label{expan} \end{eqnarray}
where $\gamma_e\simeq0.5772$ is the Euler-Mascheroni constant and
${\rm Li}^{(l_1,l_2)}(n,x)$ denotes the $l_1$-th and $l_2$-th partial
derivative of the polylogarithmic function ${\rm Li}(n,x)$
with respect to $n$ and $x$ respectively. For $\mu=0$ 
we have ${\rm Li}^{(1,0)}(0,-1)=-\ln(\pi/2)/2$ and we recover 
the known expressions of ref. \cite{Dolan:qd}.

Making use of eqs. (\ref{init}), (\ref{expan}) we can 
write the potential $U_\Lx$ in the form of eq. (\ref{mpot})
with\footnote{The quartic coupling of the
effective three-dimensional theory is given by $\lx T$.
We do not refer to this coupling, as its additional
temperature dependence is eliminated trivially
when we discuss the phase diagram in terms of the observables of
the four-dimensional theory.}
\begin{eqnarray}
m^2&=&-\lx_{0R}\, \rho_{0R}
+\left[ \frac{\lx_{0R}}{4} 
\left(1 -\frac{3}{\pi^{3/2}}  \right) 
+\frac{h^2N_c}{12}\right] T^2
+\frac{h^2N_c}{4\pi^2} \mu^2
\label{minit} \\
\lx&=&\lx_{0R}+
\frac{h^4 N_c}{16\pi^2}
\Biggl[
-1+2\gamma_e+\ln\left(\frac{\Lt^2}{4T^2}\right) \Biggr.
+ 2\,{\rm Li}^{(1,0)}\left(0,-\exp\left(\frac{\mu}{T} \right) \right)
\Biggl. 
\nonumber \\
&&~~~~~~~~~~~~~~~~~~~~~~~~~~~~~~~~~~~~~~~~~~~~~~~~~~~~~~
+2\,{\rm Li}^{(1,0)}\left(0,-\exp\left(-\frac{\mu}{T} \right) \right)
\Biggr].
\label{linit}
\end{eqnarray}
The function 
${\rm Li}^{(1,0)}\left(0,-\exp(x)  \right)+
{\rm Li}^{(1,0)}\left(0,-\exp(-x)  \right)$
is monotonically decreasing, and it takes the value 
$-\ln(\pi/2)$ for $x=0$.
These expressions demonstrate how the phase diagram that we discussed
in the previous section can emerge in the context of QCD. 
Even if the mass term is
negative at $T=\mu=0$, it can become positive for sufficiently large 
$T$ or $\mu$. For constant $T$,
the quartic coupling is a decreasing function of $\mu$. For small $h$
the tricritical
point is expected to appear for values of $T$ and $\mu$ such that 
$\lx \simeq 0$.

\section{The equation of state near the tricritical point}

\subsection{The crossover behaviour}

\begin{figure}[t]
 \centerline{\epsfig{figure=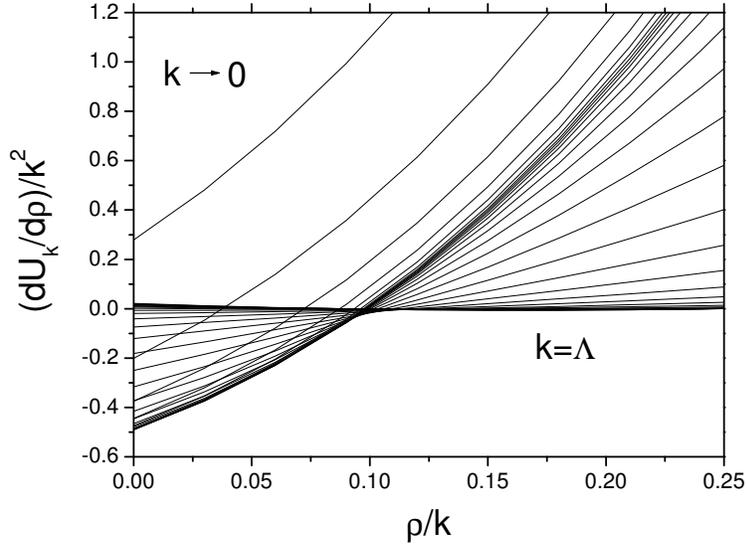,width=11cm,angle=0}}
 \caption{\it
The evolution of the rescaled potential for 
$m^2=0.0213$, $\lx=-0.335$, $g=1$.
}
 \label{fig1}
 \end{figure}

The phase structure in the absence of an external source ($j=0$)
can be determined in complete analogy to ref. \cite{Tetradis:2003qa}.
There is a line of second-order phase transitions which can be approached
for fixed positive values of $\lx$, $g$ in eq. (\ref{mpot})
by fine-tuning the mass term. During the renormalization-group flow 
the potential approaches a form characteristic of the 
fixed point of the $O(4)$ universality class. Subsequently 
the system deviates towards the broken or symmetric phase. 
The influence of the fixed point results in a behaviour that can
be characterized by universal quantities, such as critical exponents. 
For sufficiently negative values of $\lx$ the phase transitions become
first-order. Again, an appropriate fine-tuning of the mass term leads to
an effective potential with two coexisting phases. No fixed point is
approached during the flow. 

The lines of first- and second-order phase transitions 
meet at a point characterized by specific values of $m^2$ and $\lx$.
The evolution of the potential in the proximity of this point is
depicted in fig. \ref{fig1}. The initial condition is given by 
eq. (\ref{mpot}).
For the plot we integrated eq. (\ref{flow}) with $\eta=0$ (and, therefore,
$Z_k=1$). We have used a smooth infrared cutoff in the propagator
$P_k$ of eq. (\ref{fourfive}), similarly to ref. \cite{Tetradis:2003qa}.
The anomalous dimension is very small ($\eta \simeq 0.03$) in this
model and setting it equal to zero gives a good approximation to
the exact solution. 
In fig. \ref{fig1} we plot the quantity 
$u'_k/\sxt$ as a function of $\sxt^2/2$ for $\pi_i=0$ and decreasing values of
the scale $k$. In an alternative notation, that makes the $O(4)$
symmetry apparent, the figure depicts 
$(dU_k/d\rho)/k^2$ as a function of $\rht=\rho/k$, with $\rho$ defined below
eq. (\ref{mpot}). 

\begin{figure}[t]
 \centerline{\epsfig{figure=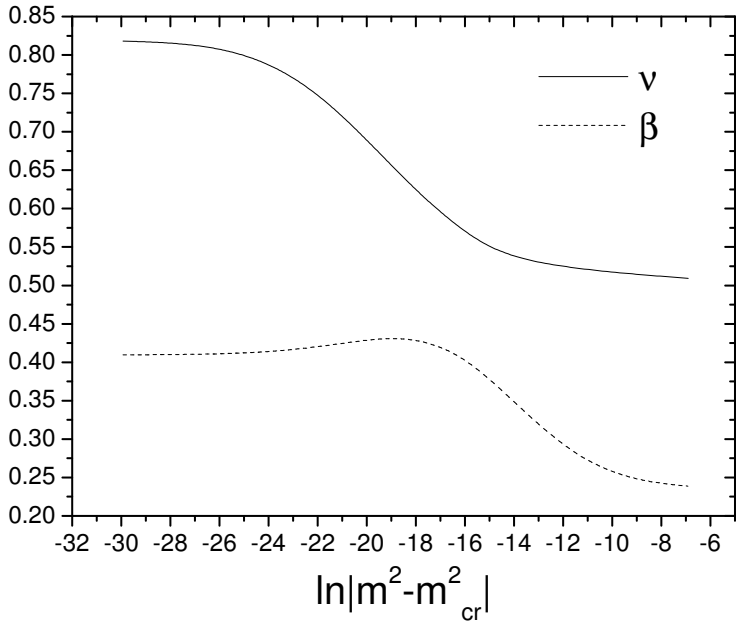,width=11cm,angle=0}}
 \caption{\it
The effective critical exponents $\nu$ and $\beta$.
}
 \label{fig2}
 \end{figure}

Two quantities must be fine-tuned for the type of evolution shown in
fig. \ref{fig1}. We
fixed $g=1$ and varied $\lx$. 
(All quantities are given in units of the ultraviolet scale $\Lx$.)
For every value 
of $\lx$ we fine-tuned $m^2$ so as to be very close to the phase transition.
For $\lx=-0.335$, $m^2_{cr}\simeq 0.0213$ (the latter 
quantity is determined with a precision of
15 significant figures) we are in the immediate vicinity of the tricritical
point and very close to a second-order phase transition.

Initially the potential is very flat in the range of $\sxt$ we are 
considering, so that its derivative is almost zero. The system is
very close to the Gaussian fixed point. 
As $k$ becomes smaller another fixed point is approached,
characteristic of the $O(4)$ universality class. Eventually the
system moves towards the symmetric phase ($u'_k/\sxt$ becomes
positive at the origin). In the limit $k\to 0$ we obtain the
effective potential, from which we can derive 
renormalized quantities such as the location of
the vacuum and the field mass.

\begin{figure}[t]
 \centerline{\epsfig{figure=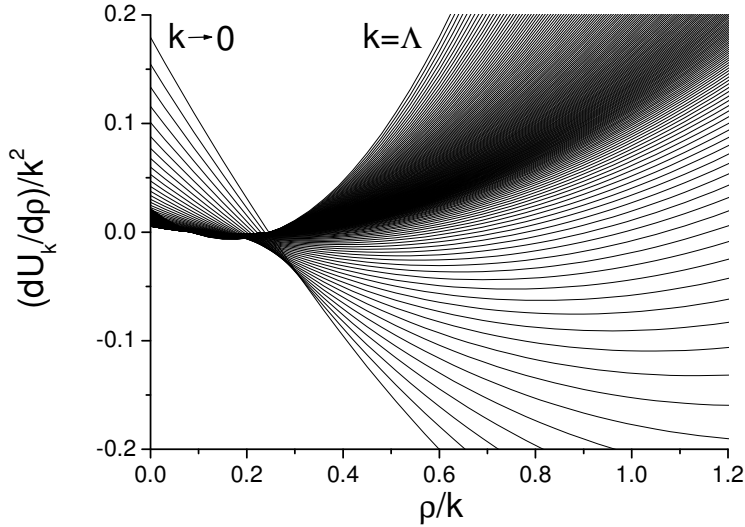,width=11cm,angle=0}}
 \caption{\it
The evolution of the rescaled potential for 
$m^2=0.0217$, $\lx=-0.338$, $g=1$.
}
 \label{fig3}
 \end{figure}

For systems close to a second-order phase transition, we can parametrize
physical quantities in terms of critical exponents. In our example,
we can keep $\lx$ and $g$ fixed with the values 
$\lx=-0.335$, $g=1$ and vary the mass term according to 
$m_R^2=m^2_{cr}+\dms$.
For $\dms>0$ the renormalized mass term at the origin can be parametrized as 
$m^2_R\sim(\dms)^{2\nu}$.
For $\dms<0$ the minimum of the potential is located at a non-zero field
value that can be parametrized as $\sx_{0R} \sim |\dms|^\beta$. 
(The quantity $\dms$ replaces the quantity $T-\tcr$ that appears when
the phase transition is discussed in the context of the
fundamental four-dimensional theory at non-zero temperature.)

The effective critical exponents $\nu$, $\beta$
are depicted in fig. \ref{fig2} as a function of $\dms$.
We observe that for relatively large $\dms$
they take the values $\beta=0.25$, $\nu=0.5$. These are the mean-field
predictions that we derived in section 2. They 
are the expected values of these exponents, as predicted by the general
theory of tricritical points \cite{lawrie}.
For large $\dms$
the potential never approaches the $O(4)$ fixed point. It stays close to
the Gaussian one, before moving directly towards the broken or the symmetric
phase. For smaller $\dms$
system feels the attraction of the $O(4)$ fixed point. For $\dms\to 0$
the potential 
spends a large part of its evolution near this
fixed point. As a result it loses memory of its initial form near the
Gaussian fixed point. The critical exponents take the values 
characteristic of the $O(4)$ universality class
$\beta=0.41$, $\nu=0.82$.\footnote{The most accurate known values for these 
exponents are close to $\beta=0.39$, $\nu=0.75$ \cite{Pelissetto:2000ek}.
The values we are quoting agree with the results of an analysis based on the
exact renormalization group within the limited truncation we are employing
\cite{Berges:2000ew}.} 
The curves of fig. \ref{fig2}
are typical examples of crossover curves. They describe the variation of
universal quantities, such as the critical exponents, as the relative 
influence of two fixed points changes.

In fig. \ref{fig3} we display the evolution of $u'_k/\sxt$ for a theory
with $m^2\simeq 0.0217$, $\lx=-0.338$, $\nu=1$.
The slight decrease of the value of $\lx$ relative to fig. \ref{fig2}
modifies drastically the evolution. Initially the potential stays very close
to the Gaussian fixed point,
but eventually moves away from it. 
Its final form is characteristic
of a first-order phase transition as was discussed in detail in ref. 
\cite{Tetradis:2003qa}.
This transition is very weak, as the discontinuity in the
order parameter is several orders of magnitude smaller than $\Lx$.

The tricritical point is located 
on the plane ($m^2$, $\lx$)
in the region between the points (0.0213, $-0.335$) and (0.0217, $-0.338$) for
$g=1$. 
For the critical values $m_t^2$ and $\lx_t$ the solution of the evolution
equation stays close to the Gaussian fixed point for an infinitely long 
``time'' $t$. As a result the critical behaviour is determined completely 
by mean-field theory. 
For example, the critical exponent $\beta$ stays equal to 0.25 arbitrarily
close to the phase transition.
The relative influence of the two fixed points depends on the 
magnitude of two parameters that can be taken as $\dms$ and $\dlx=\lx-\lx_t$.
For fixed $\dlx$ and $\dms\to 0$ the $O(4)$ fixed point dominates,
while for fixed $\dms$ and $\dlx \to 0$ only the Gaussian fixed point 
is approached. All this is consistent with the general theory of
tricritical points \cite{lawrie}.

\subsection{The universal equation of state}

All the physical information near a second-order phase transition
can be encoded in the equation of state. This describes the 
relation between the order parameter (in our case the field expectation
value), the parameters that control the distance from
the phase transition (such as $\dms$)  and the external source. 
In the case of one relevant parameter the equation of state can be
cast in the form \cite{widom,Rajagopal:1992qz} 
\begin{equation}
\delta j=\frac{dU}{d\sx}=\sx |\sx|^{\delta-1} f(x),~~~~~~~~~~~~~~~
x=\frac{\dms}{|\sx|^{1/\beta}},
\label{eos1} \end{equation}
with $U(\sx)$ the effective potential. For sufficiently small $\dms$, so
that the $O(4)$ fixed point is approached, the function $f(x)$ takes a
universal form $f_{O(4)}(x)$ 
that has been computed in ref. \cite{Berges:1997eu}.
The critical exponents $\delta$, $\beta$, as well as the other critical
exponents and
amplitudes encoded in $f(x)$, take values characteristic of this 
universality class: $\delta\simeq 4.8$, $\beta \simeq 0.39$.
(Our approximate calculation above gave $\beta \simeq 0.41$).

For a theory that only approaches the Gaussian fixed point the 
equation of state can be inferred more easily. As we saw earlier, 
because of the smallness of the various couplings near the Gaussian
fixed point, the mean-field description is sufficient. 
For a more detailed understanding of the renormalization-group flow,
we assume that the potential $U_k$ can be approximated
by a polynomial, similarly to eq. (\ref{mpot})
\be
U_k(\rht)=m^2(k)\rho+\frac{1}{2}\lx(k)\rho^2+\frac{1}{3}g(k)\rho^3
+\frac{1}{4} q(k)\rho^4 + ...
\label{mpotr} \ee
Starting from the flow equation (\ref{flow}) we can derive 
evolution equations for the various parameters 
\cite{Tetradis:1993ts}
\begin{eqnarray} 
\frac{dm^2(k)}{dk}&=&-\frac{3}{2\pi^2} \lx
\,\, l^3_1\left( \frac{m^2}{k^2} \right)
\label{ev1} \\
\frac{d\lx(k)}{dk}&=&\frac{3}{\pi^2} \frac{\lx^2}{k^2}
\,\, l^3_2\left( \frac{m^2}{k^2} \right)
-\frac{4}{\pi^2} g
\,\,l^3_1\left( \frac{m^2}{k^2} \right)
\label{ev2} \\
\frac{dg(k)}{dk}&=&-\frac{15}{2\pi^2} \frac{\lx^3}{k^4}
\,\, l^3_3\left( \frac{m^2}{k^2} \right)
+\frac{27}{2 \pi^2} \frac{\lx g}{k^2}
\,\,l^3_2\left( \frac{m^2}{k^2} \right)
-\frac{15}{\pi^2}\, q
\,\,l^3_1\left( \frac{m^2}{k^2} \right).
\label{ev3}
\end{eqnarray}
(The anomalous dimension is zero near the Gaussian fixed point.)
The most important terms in the above equations are the term in
the r.h.s. of the first equation and the last term in the r.h.s. 
of the second equation.
These generate contributions to the renormalization of $m^2$ and
$\lx$ that are proportional to the ultraviolet scale $\Lx$.
The last term in the r.h.s. of the third equation does not
generate significant contributions because $h=0$ in the bare potential
of eq. (\ref{mpot}).

For a potential that is given by eq. (\ref{mpot}) for $k=\Lx$,
and a flow that stays in the vicinity of the 
Gaussian fixed point, the 
integration of the evolution equations is expected to renormalize
significantly only 
$m^2$ and $\lx$. The coupling $g$ is expected to receive only logarithmic
corrections (it corresponds to a marginal operator)
cut off by the mass term.\footnote{It 
must be pointed out that
no significant infrared effects are expected near the Gaussian
fixed point because of the decoupling of the various modes at low
enough scales. This happens in the symmetric phase, around
which we expanded the potential in eq. (\ref{mpotr}). However, if
the vacuum is at a non-zero value of the field the contributions
of the three massless pions (the Goldstone modes) never decouple. 
This leads to interesting effects that have been discussed in 
ref. \cite{Tetradis:1992xd}, for example.
} Moreover, these corrections are expected to
reduce $g(k)$ and shift the potential closer to the Gaussian fixed point
\cite{drouffe}. 
These conclusions are verified by the numerical integration of 
the full evolution equation (\ref{flow}).
 
If we neglect the logarithmic corrections
a good approximation of the effective potential is 
\be
U(\rho)=m^2_R\rho+\frac{1}{2}\lx_R\rho^2+\frac{1}{3}g\rho^3.
\label{mpotrr} \ee
The tricritical point corresponds to the theory with $\lx_R=0$, for
which the equation of state can be written as
\begin{equation}
\frac{4j}{g}=\sx^5 \left( 1+
\frac{4m^2_R/g}{\sx^4}\right).
\label{eqstattri} 
\end{equation}
Comparison with eq. (\ref{eos1}) indicates that for the tricritical
point $\dms=4m^2_R/g$, $\delta j=4j/g$, 
$\delta=5$, $\beta=1/4$, $f_{tr}(x)=1+x$. 
This is in agreement with the values of the exponents in fig. \ref{fig2} for
relatively large $\dms$.

For $\lx_R\not= 0$ the equation of state can be written in terms of
the potential of eq. (\ref{mpotrr}) with the identification
$\dms=m^2_R$, $\dlx=\lx_R$. We expect a crossover away from the
tricritical point for $\dlx > \dms$. This is apparent in fig. (\ref{fig2})
in the transition of the effective exponent $\beta$ from
the tricritical-point value $\beta=0.25$ to a value close to 0.5.
Of course, the asymptotic value is determined by the $O(4)$ fixed point that
attracts the flows after they leave the Gaussian fixed point.
In analogy with the crossover observed in fig. \ref{fig2} 
for the critical exponents,
we expect the critical equation of state to change continuously between the
form characteristic of the tricritical point  and the one of
the $O(4)$ universality class, as $\lx$ is varied.

\section{The equation of state near the critical point}

\subsection{The vicinity of the tricritical point}

As we have seen in section 3, for non-zero external source $j$ and
negative $\lx$ it is possible to have a second-order phase transition at
the end of a line of first-order phase transitions.
The order parameter is the deviation of the field from a certain non-zero 
expectation value. This type of transition is not associated with the breaking
of a symmetry, as the original symmetry is explicitly broken by the
source. 

For the part of the line of 
critical points in the vicinity of the tricritical point the potential is
approximately given by eq. (\ref{mpotrr}) 
(neglecting logarithmic corrections) and the analysis of section 3 based on
mean-field theory is sufficient. 
As we saw there, 
we must define a new effective potential
$\Ut(\sx;j)=U(\sx,\pi_i=0)-j_* \sx$ whose minimum is located away from 
the origin. The value of the source $j_*$ must be defined appropriately.

In order to derive the
equation of state we shift the field:
$\sx \to \sx_*+\sx$. The value of $\sx_*$ can be determined by
requiring that at the critical point the minimum of the potential is
located at $\sx=0$. As we saw in section 3, at the critical point we have 
$d\Ut/d\sx=d^2\Ut/d\sx^2=d^3\Ut/d\sx^3=0$ at the minimum. 
The last equality leads to $\sx^2_*=-3\lx_R/(5g)$, indicating that
the existence of a critical point requires $\lx_R < 0$. 

On a surface of
constant $\lx_R$ and $\sx_*$ (we assume a fixed renormalized $g$ 
throughout the paper) the effective potential for the shifted field
can be written as 
\begin{equation}
j=\frac{dU}{d\sx}=
\frac{2}{3} g \sx_*^5
+\sx_*\dmst+\dmst\sx+\frac{5}{3}g \sx^2_* \sx^3
+\frac{5}{4} g \sx_* \sx^4+\frac{1}{4} g \sx^5,
\label{eosg} \end{equation}
where we have 
redefined the mass term 
according to $m^2 = m^2_*+ \dmst$, with
$m^2_*=5g\sx_*^4/4$.
If we repeat the calculation for a different value of $\lx_R$, the form
of the above equation remains the same. The only parameter that
changes is $\sx_*$ through its dependence on $\lx_R$ given above. 
The field $\sx$ corresponds now 
to deviations from the new value of $\sx_*$. 

It is convenient to define $\sx_*=\sx_{*0}-\ex$, with $\sx_{*0}$ being
kept fixed and $\ex$ accounding for variations of $\lx_R$ around a 
constant value. Moreover, the field $\sx$ can be 
redefined as the deviation from
$\sx_{*0}$. Then, eq. (\ref{eosg}) retains its form with the replacements
$\sx_* \to \sx_{*0} -\ex$, $\sx \to \sx +\ex$, and can be written as
\begin{equation}
\dj=F_1(\ex,\dms)+
(\sx+\ex)^3\left(\frac{\dms}{(\sx+\ex)^2}+1 \right)
+{\cal O}\left( (\sx+\ex)^4\right),
\label{eosg1} \end{equation}
where $j=j_*+\djt$, $j_*=2g \sx_{*0}^5/3$,
and 
$\dj=3\, \djt/(5g\sx_{*0}^2)$, $\dms=3\, \dmst/(5g\sx_{*0}^2)$.
We assume that $|\sx|$ takes values in a range of a few $|\ex|\ll \sx_{*0}$.
The function $F_1(\ex,\dms)$ is zero for $\ex=\dms=0$. Near the
tricritical point it
takes the form
$F_1(\ex,\dms)\simeq -2\sx_{*0}^2\ex +4\sx_{*0}\ex^2 -4\ex^3+
\sx_{*0}\dms$.

It must be emphasized that the above expression is a rather complicated way
to rewrite the potential of eq. (\ref{mpotrr}). As a result, the phase
diagram as a function of $\dms$, $\dj$ and $\ex$ is expected to
have the same structure as the 
one discussed in section 3. This can be verified easily.
There is a line of critical points parametrized by $\ex$ for
$\dms=\dj=0$. A particular point 
can be approached on a surface of constant $\ex$. 
For example, on the surface $\ex=0$ the critical 
point can be approached along the
line $\dj=F_1(0,\dms)=\sx_{*0}\, \dms$. 
The resulting second-order
phase transition is parametrized by the critical exponents 
of mean-field theory: $\beta=1/2$, $\gamma=1$.
For $\ex=0$, $\dms<0$ the potential has two minima and 
there are two possible vacuum states. A first-order 
transition from the vicinity of one vacuum
to the other takes place when $\dj$ is varied through 
$\sx_{*0}\, \dms$. For $\dms>0$ the transition from positive
to negative field expectation 
values is continuous, but the
mass term never becomes zero. This situation corresponds to an analytical
crossover.
The qualitative picture remains the same for other values of $\ex$.

The term $F_1(\ex,\dms)$ in eq. (\ref{eosg1}) can be absorbed in 
the source term $\dj$ through a redefinition of the potential. However,
our choice of $\dms$, $\dj$ gives the most transparent 
connection with the parameters used for the phase diagram of QCD.
The term $\dj$ corresponds to deviations of the explicit symmetry-breaking
term in the bare Lagrangian from some constant value. In this sense
it corresponds to variations of the current quark mass.
The terms $\ex$, $\dms$ are related to parameters of the theory that
do not break explicitly the original symmetry. In the QCD case, 
they are functions of the temperature and the baryonic chemical potential. 

Within the quark-meson model, that constitutes an effective description
of low-energy QCD, the parameters $\ex$, $\dms$ can be expressed as
functions of $\dt$ and 
$\dm$, where $\dt=T-T_*$, $\dm=\mu-\mu_*$ are the deviations from the values
that correspond to the 
second-order phase transition for a given current quark mass.
We can integrate eqs. (\ref{ev1})--(\ref{ev3}) from $k=\Lx=T$ down to
$k=0$ 
using (\ref{minit}), (\ref{linit}) as initial conditions for
$m^2$ and $\lx$, and taking 
$g={\cal O}(h^6)\simeq 0$.\footnote{Here $h$ is the 
Yukawa coupling that determines the constituent quark mass. We assume that it
is small, consistently with the derivation of 
eqs. (\ref{minit}), (\ref{linit}).}
The only significant contribution resulting from this integration is
a term 
$3\lx_{0R}T^2/(4\pi^{3/2})$ that must be added to
the mass term $m^2$ \cite{Tetradis:2003qa}.
We find
\begin{eqnarray}
\dms&\sim&
\left( \lx_{0R}
+\frac{h^2N_c}{3}\right) T_* \dt
+\frac{h^2N_c}{\pi^2} \mu_* \dm
\label{dminit} \\
\ex&\sim&\frac{1}{T_*}\dt
+ {\rm Li}^{(1,1)}\left(0,-\exp\left(\frac{\mu_*}{T_*} \right) \right)
\exp\left(\frac{\mu_*}{T_*} \right)
\left(\frac{1}{\mu_*}\dm-\frac{\mu_*}{T^2_*}\dt
\right)
\nonumber \\
&&~~~~~~~~
+ {\rm Li}^{(1,1)}\left(0,-\exp\left(-\frac{\mu_*}{T_*} \right) \right)
\exp\left(-\frac{\mu_*}{T_*} \right)
\left(-\frac{1}{\mu_*}\dm+\frac{\mu_*}{T^2_*}\dt
\right).
\label{dlinit}
\end{eqnarray}

In the following subsection we shall see that it is possible to cast
the equation of state in a form similar to eq. (\ref{eosg1}) even away from the
tricritical point.
An important observation is that there are
many ways to approach a critical point. The universal properties of the
physical system (such as the values of the 
critical exponents) are not the same along every path.
Contrary to the case discussed above, the critical exponents can take effective
values different from the expected ones. For example, if the critical point
is approached on the surface $\dj=0$, along a line  
$\ex=\ex(\dms)$ such that $F_1(\ex(\dms),\dms)= 0$, 
we obtain $\sx+\ex(\dms)\sim |\dms|^{1/2}$ for the location of the minimum of
the potential when $\dms<0$.
The mass term at the vacuum scales $\sim |\dms|$, and the exponent
$\gamma$ takes the value $\gamma=1$.
If the critical point is approached along a line 
$\ex=\ex(\dms)$ such that $F_1(\ex(\dms),\dms)\not= 0$, 
we obtain $\sx+\ex(\dms)\sim |\dms|^{1/3}$ for the location of the minimum of
the potential when $\dms<0$.
The mass term at the vacuum scales $\sim |\dms|^{2/3}$, and the exponent
$\gamma$ takes an effective value $\gamma_{eff}=2/3$.
In the case of the quark-meson model,
it is obvious from the form of eqs. (\ref{dminit}), (\ref{dlinit})
that the generic trajectory that passes through the critical point 
has $F_1(\ex(\dms),\dms)\not= 0$ with high probability.

These findings are consistent with the general analysis of critical behaviour
in the vicinity of tricritical points \cite{lawrie}. 
According to this analysis one must introduce more than one scaling fields 
\cite{riedel,chang,fisher}, and the critical behaviour depends on their
relative magnitude.
We emphasize, however, that 
the values that we derived above do not correspond to a tricritical
theory. The reason is that we are studying the theory near the line of
critical points. The part of this line that lies in the vicinity of
the tricritical point is described by mean field theory. As a result 
we find the mean-field exponents of the critical and not the tricritical
theory.

\subsection{Away from the tricritical point}

Far from the Gaussian fixed point the theory is renormalized significantly and
the potential cannot be approximated by a simple polynomial any more.
The flow equation (\ref{flow}) describes the evolution of the potential
as the scale $k$ is lowered from $\Lx$ to zero. The last term in the
r.h.s. of this equation includes contributions from the radial mode
(the $\sx$ field) and the three Goldstone modes (the pions). 
In the presence of an external source the ground state of the system
is located at a point $\sx_*$ 
where the potential has a non-zero derivative and
satisfies $dU/d\sx=j_*$. For zero anomalous dimension,
the argument of the 
second ``threshold''
function is expected to become asymptotically 
$u'_k/\sxt \to (j_*/\sx_*)/k^2$ for $k\to 0$. 
As a result the pions are expected to 
decouple and their contribution to the evolution to switch off.
(A small anomalous dimension does not affect this conclusion.)
This decoupling behaviour has been demonstrated explicitly in ref. 
\cite{Tetradis:2003qa}, and we shall not repeat the analysis here.

Deep in the infrared, after the pion decoupling, the flow equation 
takes the form 
\begin{equation}
\frac{\partial}{\partial t} u_k(\sxt)=-3 u_k + \frac{1}{2}(1+\eta)
\, \sxt  u'_k+ \frac{1}{4\pi^2}\,\, l^3_0\left( u''_k\right).
\label{flowd} \end{equation}
Apart from the Gaussian fixed point, the
only known $t$-independent solution of this equation is the Ising
fixed point, for which 
$\partial u_I/\partial t =0$ \cite{Morris:1996kn}.
This solution has a $Z_2$ symmetry $\sx \leftrightarrow -\sx$.
Whether the fixed point will be approached depends on the initial
condition for the potential. If the potential at some scale $k=\Lt$ after
pion decoupling is
$Z_2$-symmetric, the subsequent flow preserves the symmetry. One
needs to fine-tune only one parameter, the mass term, in order to
approach the Ising fixed point. The theory 
has only one relevant parameter. 

However, in the presence of a source the potential
$\Ut_{\Lt}(\sx;j)=U_{\Lt} 
(\sx,\pi_i=0)-j_*\sx$ does not possess a $Z_2$ symmetry and
the flow becomes more complicated. 
If we expand $\Ut_k$ after shifting the
field by a $k$-independent value $\sx\to \sx_*+\sx$, we 
obtain a polynomial with several even and odd powers of $\sx$.
The crucial question is how many of them need to be fine-tuned in 
order to approach the Ising fixed point.

In order to answer this question we must study the small perturbations 
around the Ising fixed-point solution $u_I$ and examine how many have negative
eigenvalues so that they grow in the infrared. In linear perturbation theory
around the fixed point ($u_k=u_I+\du_k$) the evolution equation becomes
\begin{equation}
\frac{\partial}{\partial t} \du_k(\sxt)=-3 \, \du_k + \frac{1}{2}(1+\eta)
\sxt \, \du'_k-
\frac{1}{4\pi^2} \, \du''_k \,
l^3_1\left( u''_I\right),
\label{flowl} \end{equation}
where we have used the property $l^d_1(w)=-d[l^d_0(w)]/dw$ of the 
``threshold'' functions defined in eq. (\ref{fourfive}).

There is one even solution $\du_{1k}$
of the above equation,  whose eigenvalue $\lx_1$ is related
to the exponent $\nu$ in the Ising universality class through $\lx_1=1/\nu$. 
The simplest odd perturbation is
$\du_{2k}=c_2\, \sxt$, with an eigenvalue $\lx_2=-(5-\eta)/2$.
There is one more odd perturbation with negative eigenvalue.
It is given by $\du_{3k}=c_3\, \partial u_I/\partial\sxt$ 
and its eigenvalue is $\lx_3=-(1+\eta)/2$
\cite{Pelissetto:2000ek,Tsypin:2001ix}.
This perturbation  corresponds to the symmetry under field shifts
of the evolution equation before the rescalings by $k$ in eqs. (\ref{scalvar})
\cite{Morris:1994jc}. It is not expected to generate genuinely new physical
behaviour, as it corresponds to a redundant operator \cite{wegner}.
The dominant non-trivial behaviour near the Ising fixed point is associated 
with the even perturbation. The two odd perturbations we discussed
are related either to a
shift of the source term (that is linear in $\sx$) or a field shift.
All other perturbations have positive eigenvalues and are expected to
become negligible in the infrared.

In spite of the apparent absence of strong effects associated
with the odd perturbations, the number of parameters that need to
be fine-tuned in order to approach the fixed point is three in the absence
of a $Z_2$ symmetry. In the model we are considering, with a bare potential
given by eq. (\ref{mpot}), we have three parameters at our disposal:
$m^2$, $\lx$, $j$. Moreover, the field value $\sx_*$ around which we expand
the potential is not determined {\it a priori}. If we fix it arbitrarily, we
need to fine-tune all three of $m^2$, $\lx$, $j$ 
in order to approach the Ising fixed point.
Equivalently, 
for an arbitrary value of $j$ the fixed point can be approached
by fine-tuning $m^2$, $\lx$, $\sx_*$. We have verified this conclusion
through the numerical integration of the evolution equation.
If the three-dimensional theory results from
a more fundamental theory related to QCD, such as the quark-meson model,
the parameters in the potential
are a function of the temperature $T$ and the chemical potential $\mu$, 
while the source term $j$ is proportional to the current quark mass.
Then we conclude that for given $j$ the fine-tuning of $T$ and $\mu$
can lead to a second-order phase transition. By varying $j$ a line 
of second-order phase transitions appears in the phase diagram.

We are interested in the universal equation of state near the phase transition.
For this reason we need to discuss the evolution of the potential
after the fixed point has been approached. The fine-tuning of the
bare parameters guarantees that the system will remain near the fixed point
for a long ``time'' $t$. However, if the fine-tuning is not perfect, 
the perturbations with negative eigenvalues will start growing eventually.
We can solve the evolution equation (\ref{flowd}) if we split $u_k$ into
even and odd parts as
$u_k=u_k^s+u_k^a$ and assume $u^a_k\ll u^s_k$. This assumption is a good 
approximation, especially in the vicinity of the fixed point,
because the symmetric part includes the fixed-point contribution $u_I$.
The even part satisfies eq. (\ref{flowd}), while for the odd part
we use the ansatz $u^a_k(\sxt)=f_2(t)\, \sxt+f_3(t)\, u^{s\prime}_k(\sxt)$.
This leads to the equations
\begin{eqnarray}
\frac{df_2}{dt}&=&-\frac{1}{2}(5-\eta)f_2,
\label{f2} \\
\frac{df_3}{dt}&=&-\frac{1}{2}(1+\eta) f_3
\label{f3} 
\end{eqnarray}
whose solution is $f_2=c_2 k^{-5/2}Z_k^{-1/2}$, 
$f_3=c_3 k^{-1/2}Z_k^{1/2}$, with $Z_k$ the wavefunction renormalization
and $c_2$, $c_3$ arbitrary constants.
This solution is consistent with our previous discussion of the perturbations
near the fixed point (where $\eta$ is constant), but also describes the
final evolution of the potential away from it.

For $k\to 0$ we obtain for the effective potential
\begin{equation}
\Ut(\sx)= U(\sx)-j_* \sx=U^s(\sx)+c_2\, \sx + c_3 \, U^{s\prime}(\sx) \simeq
U^s(\sx+c_3)+c_2\, \sx.
\label{eosa} \end{equation}
We know that $U^s(\sx)$ is related to the universal equation of state of
the Ising model through eq. (\ref{eos1}). This implies that we
can parametrize the equation of state similarly to eq. (\ref{eosg1})
\begin{equation}
\dj=\frac{d\Ut}{d\sx}=F_2(\ex,\dms) 
+ (\sx+\ex)|\sx+\ex|^{\delta-1} f_{Z_2}(x),
~~~~~~~~~~x=\frac{\dms}{|\sx+\ex|^{1/\beta}},
\label{eos2} \end{equation}
where $\dms,\ex \to 0$ and 
$|\sx|$ takes values in a range of a few $|\ex|$.
The universal
function $f_{Z_2}(x)$ is specified by the Ising universality class
\cite{Berges:2000ew,Pelissetto:2000ek}. The critical exponents are
$\beta= 0.33$, $\delta = 4.8$.
For an effective theory resulting from QCD,
the parameters $\dms$, $\ex$ are functions of the temperature and 
chemical potential, while $\dj$ corresponds to deviations of the quark mass
from a constant value.

The appearance of the function $F_2(\ex,\dms)$ can be understood as follows:
For a potential expanded around a fixed value $\sx_{*0}$,
the parameters $c_3$ (related to $\ex$), $\dms$ (used in the parametrization
of the symmetric part) are functions of the parameters $m^2$, $\lx$
of the bare theory (or $T$, $\mu$ for QCD). The parameter $c_2$ depends on
$j_*$ as well, but in a trivial way: $c_2=-j_*+G(m^2,\lx)$. 
The choice of $j_*$ does not
affect the renormalization flow because 
the evolution equation involves second functional 
derivatives with respect to the field.
As a result, a term linear in $\sx$ does not change through the evolution and
can be added directly to the effective potential. 
On the other hand, if the bare potential is not $Z_2$-symmetric its first
derivative at the origin changes during the evolution
and can become non-zero, even if it was zero for the bare theory.
Expressing $m^2$, $\lx$ in terms of $\dms$, $\ex$ leads to the expression
$c_2=-j_*+F_2(\ex,\dms)$. The fine-tuning of $j_*$ amounts to
choosing it so that it cancels $F_2(0,0)$. This leads to eq. (\ref{eos2}),
where $F_2(0,0)=0$ and $\dj$ measures deviations of the source from $j_*$.
The function $F_2(\ex,\dms)$ can be determined
for a given bare theory similarly to the 
case of eq. (\ref{eosg1}). A simple analytical calculation is not 
feasible, as perturbation theory is not applicable away from
the tricritical point. The evolution equation has to be integrated 
numerically while 
performing a triple fine-tuning of the initial conditions
so that the Ising fixed point is approached.

As expected the odd perturbations do not lead to the introduction of
a new universal function. They simply generate a field shift. 
Our result is in agreement with ref. \cite{Seide:1998ir}.
In that study the bare potential was assumed to have a very simple form
(a polynomial of fourth degree). Our result provides the generalization
to the case of an arbitrary bare potential.

Our conclusions are not expected to be altered when higher terms in 
the derivative expansion are taken into account. The form of the odd
perturbations will remain the same. The first perturbation corresponds to
a term linear in the field. The introduction of
such a term does not modify the exact flow
equation for the effective action \cite{Wetterich:yh,Berges:2000ew} before
the rescaling by $k$ in eqs. (\ref{scalvar}), as this equation
involves second functional derivatives with respect to the field.
As a result the implications of such a term are independent of the truncation.
Similarly, the second odd perturbation is related to a redundant operator
that leads to field shifts. Again, 
its effects are expected to be independent of
the truncation.

In analogy with the crossover for the equation of
state between the form characteristic of the
tricritical point (eq. (\ref{eqstattri})) and that of the $O(4)$ universality
class (eq. (\ref{eos1}))
in the case with $j=0$, we expect a different
crossover as a function of $j_*$.
For $j_* \to 0$ the universal equation of state near the critical point 
is given by eq. (\ref{eosg1}), while for large $j_*$ it takes the form
of eq. (\ref{eos2}). A crossover is expected between the two forms as
$j_*$ is varied continuously.

\section{Implications for QCD}

Because of its universal form, eq. (\ref{eos2}) is expected to 
describe the critical behaviour of all physical systems whose phase diagram
includes a critical point similar to that of our toy model.
For two-flavour QCD the connection has been made more explicit
in ref. \cite{Tetradis:2003qa}. It was shown that a potential
whose leading terms are given by eq. (\ref{mpot}) results, 
at energy scales below the temperature, from
the integration of the fermionic contributions in a model of 
quarks coupled to mesons. The subsequent decoupling of the pions at lower
scales leads to a critical theory with only one massless field,
the $\sx$ field. The universal equation of state is given by eq. (\ref{eos2}),
with $\dms$, $\ex$ proportional to deviations of the temperature $\delta T$
and the
baryonic chemical potential $\delta \mu$
from certain values. The source term $\dj$ is
proportional to the deviation of the quark mass from a constant value.

It is obvious that in experimental situations 
the critical point of QCD can be approached only along the 
surface $\dj=0$. The details of the experiment (center of mass energy,
type of colliding nuclei) determine the effective temperature and 
chemical potential. Information from various experiments can be used
in order to approach the critical 
point along a curve $\ex=\ex(\dms)$. An important
question is whether the universal properties of the critical system (such
as scaling parametrized by critical exponents) are observable.
The solution of eq. (\ref{eos2}) with $\dj=0$, $\ex=\ex(\dms)$
determines the location of the vacuum. The function $F_2(\ex(\dms),\dms)$ is a
regular function around the point (0,0) and can be Taylor expanded. (An 
explicit example is given for $F_1(\ex,\dms)$ below eq. (\ref{eosg1}).)
As $F_2(0,0)=0$ the leading term is $c\, \dms$, with $c$ some constant.
(We have assumed $\ex=\ex(\dms)$.)
The solution of eq. (\ref{eos2}) is 
$|\sx+\ex(\dms)|=|c\, \dms/D|^{1/\delta}$, where $D=f(0)$.
This solution emerges because 
the critical exponents $\delta=4.8$, $\beta=0.33$
satisfy $\beta\delta >1$.

The ``unrenormalized'' 
mass term $d^2\Ut/d\sx^2$ (equal to the inverse susceptibility)
scales as $|\sx+\ex(\dms)|^{\delta-1}
\sim |\dms|^{(\delta-1)/\delta}$. This implies that the effective exponent
$\gamma_{eff}$ takes the value $\gamma_{eff}=(\delta-1)/\delta=0.79$.
This should be compared with the standard value $\gamma=1.24$ 
in the Ising universality 
class \cite{Berges:2000ew,Pelissetto:2000ek}. 
For the exponent $\nu$ parametrizing the divergence of the correlation
length we find $\nu_{eff} =  0.40$. This
is a consequence of the scaling law $\nu=\gamma/(2-\eta)$ that relates
$\nu$, $\gamma$ and the small
anomalous dimension $\eta=0.036$. The standard value of $\nu$ in
the Ising universality class is $\nu=0.63$.

It must be pointed out that the mass term may scale with the standard value for
the exponent along certain paths that approach the critical point.
For example, for $\ex=0$, $\dj=F_2(0,\dms)$ the standard scaling is obtained.
However, this approach to the critical 
point is unphysical as the quark mass cannot
be altered. Another possible path has 
$\dj=0$ and $\ex(\dms)$ given by the
solution of the equation $F(\ex,\dms)=0$. Again, such a fine-tuned path
is unlikely to be realized experimentally. 
These conclusions are in agreement with previous
studies of the tricritical and critical points,
in which the arguments were based on scaling relations
\cite{Stephanov:1998dy}. 

The effective values for the exponents that we derived above are smaller
than the standard ones in the Ising universality class by approximately 
40\%. This implies that the divergence of quantities such as the correlation
length or the susceptibility  along
the experimentally accessible paths is much slower than the naive expectation.
This conclusion is a consequence of the
fact that the external source (the explicit symmetry breaking term) cannot
be altered, as it is proportional to the quark mass. 
The smallness of the effective exponents implies that the universal behaviour
is not easily accessible. The critical point must be approached very 
closely before the divergences of the correlation length or the susceptibility
become apparent. 

As a final comment we mention that the inclusion of the strange quark is
not expected to modify our conclusions. It affects primarily the location of
the line of critical points, while the general structure of the phase 
diagram remains the same. The general conclusion is
that there is a direction in the phase
diagram that is not under experimental control. It is related to the
explicit breaking effects of the chiral symmetry ($SU(3)_L\times SU(3)_R$
in the case of three flavours) that are associated with the masses of the
quarks. As a result, the experimental 
approach to a critical point is expected to
follow a path that displays the effective exponents we derived above.

\vspace{0.5cm}
{\bf Acknowledgements:} We would like to thank T. Morris, M. Stephanov, 
E. Vicari and C. Wetterich for many useful discussions.

\end{document}